\documentstyle[twocolumn,aps,epsfig,floats]{revtex}
\draft
\begin{document}
\twocolumn[\hsize\textwidth\columnwidth\hsize\csname @twocolumnfalse\endcsname

\title{Spintronic and electrochromic device based on
Li-intercalated transition-metal doped anatase $\rm TiO_2$}

\author {Min Sik Park and B. I. Min}

\address{Department of Physics and electron Spin Science Center,  \\
         Pohang University of Science and
	 Technology, Pohang 790-784, Korea}

\date{\today}
\maketitle

\begin{abstract}
We have explored the effects of the Li intercalation on the electronic and
magnetic properties of transition-metal (TM) doped anatase TiO$_2$.
By Li intercalation, Mn-doped TiO$_2$ exhibits the insulator to metal 
transition.
On the other hand, Li-intercalated Fe-doped TiO$_2$ has the
insulating ground state  for low concentration of Li/Ti=0.067, but 
the metallic ground state for high concentration of Li/Ti=0.133.
We discuss the $n$-type carrier induced ferromagnetism in 
Li-intercalated TM-doped anatase TiO$_2$.
Based on the Li-intercalated TM-doped anatase TiO$_2$,
we propose a potential spintronic and electrochromic device 
controlled by the electric-field.

\end{abstract} 

]


Spintronics, namely spin-based electronics, is a new generation of 
microelectronics which utilizes both charge and spin degrees of freedom of 
carriers.  Dilute magnetic semiconductors (DMSs) are expected to 
play a vital role in
spintronics due to easy integration into existing semiconductor devices.
Ideal DMS must satisfy such conditions as high Curie temperature (T$_C$) 
above room temperature and easy incorporation of $p$- and
$n$-type dopants.
So far, two types of Mn-doped DMS families have been investigated :
II-VI (CdTe and ZnSe)\cite{Furdyna} 
and III-V (GaAs)\cite{Ohno} zinc-blende compounds. 
Especially, the ferromagnetic (FM) behavior with $T_{C} \sim 110K$ of
Mn-doped GaAs attracts great attention,
and it has been suggested that delocalized 
holes mediate the FM interaction between Mn spins\cite{Dietl}. 

It has been recently reported that Co-doped anatase TiO$_2$ film 
($\rm Ti_{1-x}Co_xO_2$) shows the ferromagnetism even above room temperature.
$\rm Ti_{1-x}Co_xO_2$ films were fabricated by means of the 
pulsed-laser-deposition\cite{Matsumoto} or 
the oxygen-plasma-assisted\cite{Chambers} molecular-beam-epitaxy technique.
However, still controversial is the issue that
the ferromagnetism in this system is an intrinsic DMS property or not. 
One claim is that Co atoms substitute properly for Ti atoms 
\cite{Matsumoto,Chambers,Chambers1} and the 
ferromagnetism is caused by the exchange interaction mediated by the
vacancy-induced $n$-type carriers.  Another is that
Co atoms form nano-clusters, resulting in the high T$_C$
ferromagnet\cite{Kim1,Kim2}.
The other claim is that Co atom incorporates in the interstitial position
or forms Co-Ti-O complexes\cite{Simpson}. 
It is also reported that the as-grown film has coexisting contributions 
from Co metal clusters and matrix-incorporated Co, 
but that the high temperature heat treatment enhances 
the matrix-incorporated Co contribution drastically\cite{Shinde}.

Independently of the DMS project, Li intercalation in anatase TiO$_2$
has been studied extensively for possible uses in high energy density 
batteries,
electrochromic, and solar-cell devices\cite{Hare,Ohzuku,Regan,Cantao}.
It is possible to intercalate Li atoms in anatase TiO$_2$
 up to the Li/Ti ratio of $\sim$ 0.7.
There is an indication that the intercalation does not occur 
uniformly throughout each crystallite \cite{Luca}. Further,
there seem to exist two phases: the Li-poor anatase phase 
and the Li-rich orthorhombic phase\cite{Wagemaker,Marina}. 
The insulator to metal transition is observed for 
the Li/Ti ratio of $\sim$ 0.3 \cite{Luca}.
Interestingly, the  magnetic susceptibility is observed to be
proportional to the Li/Ti ratio in the Li-intercalated
anatase TiO$_2$ with
the measured localized moment of $\sim$ 0.004 $\mu_B$ per Ti.

Motivated by the easy Li intercalation into anatase TiO$_2$,
we have explored the magnetic properties of 
transition-metal (TM) doped TiO$_2$ with intercalating Li.
To this end, we have investigated electronic structures of 
Li-intercalated $\rm Ti_{0.9375}M_{0.0625}O_2$ (M=Mn, Fe).
We have found that both Mn- and Fe-doped TiO$_2$ with sufficient
intercalated Li concentration have metallic and ferromagnetic ground
states, implying that $n$-type carriers produced by Li intercalation 
induce the ferromagnetism.

We have employed the linearized muffin-tin orbital (LMTO) 
band method in the local-spin-density approximation (LSDA).
The space group of anatase structure is tetragonal $I4_{1}/amd$.
The anatase TiO$_2$ is composed of 
stacked edge-sharing octahedrons formed by six O anions.
Ti atoms are in the interstitial sites of octahedrons
that are distorted with different bond lengths 
between the apical (1.979 $\AA$) and 
the equatorial (1.932 $\AA$) Ti-O bond and with the Ti-O-Ti angle 156.3$^{o}$.
For $\rm Ti_{0.9375}M_{0.0625}O_{2}$, we have considered 
a supercell containing sixteen formula units in the primitive unit cell
by replacing one Ti by M atoms ($\rm Ti_{15}M_{1}O_{32}$: 
$a=b=7.570, c=9.514$ \AA).
Sixteen empty spheres are considered in the interstitial sites to enhance 
the packing ratio for the LMTO band calculation.
To simulate the Li intercalation, we replaced one (corresponds to 
Li/Ti=0.067) or two (Li/Ti=0.133) interstitial empty spheres 
by Li atoms.

\begin{figure}[t]
\centerline{
\epsfig{file=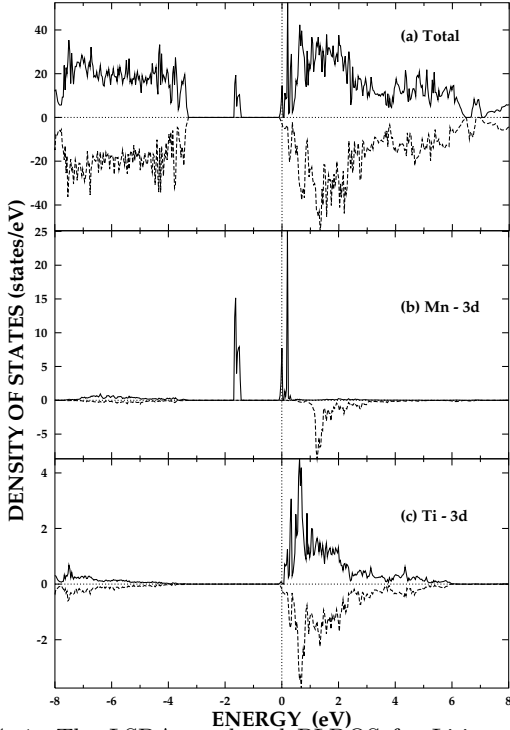,width=7.0cm}}
\caption{The LSDA total and PLDOS
   for Li-intercalated $\rm Ti_{0.9375}Mn_{0.0625}O_{2}$ (Li/Ti=0.067).
}
\label{MnLi1}
\end{figure}


We have studied before electronic structures of TM doped 
anatase $\rm Ti_{1-x}M_xO_2$ (M=Mn, Fe, Co, Ni) without
Li intercalation\cite{Mspark}.
Using the LSDA band calculation, 
we have obtained the {\it half-metallic} ground states 
for Fe- and Co-doped cases with the carrier type of
mainly Fe and Co 3$d$ states, respectively.
On the other hand, we have obtained insulating ground states 
for Mn- and Ni-doped cases.  Ni-doped TiO$_2$ was found to be nonmagnetic.
We have also studied Li intercalation effects on the
electronic and magnetic properties of both undoped and Co-doped 
anatase TiO$_2$ \cite{Mspark1}. 
What we have found for undoped TiO$_2$ was that the $n$-type carriers 
produced by the Li intercalation fill Ti 3$d$ conduction band so as to
induce small magnetic moments at Ti sites.
In contrast, Li-intercalated Co-doped TiO$_2$ becomes
nonmagnetic and insulating, because the produced $n$-type carriers 
fill up the low-spin Co 3$d$ states located in the energy gap region.
Hence the Li intercalation was detrimental to stabilizing the magnetic
state in Co-doped TiO$_2$.
These features, however, suggested that some other TM-doped anatase 
TiO$_2$ with TM of high-spin configuration would have carrier induced 
ferromagnetism through the Li intercalation.

\begin{figure}[t]
\centerline{
\epsfig{file=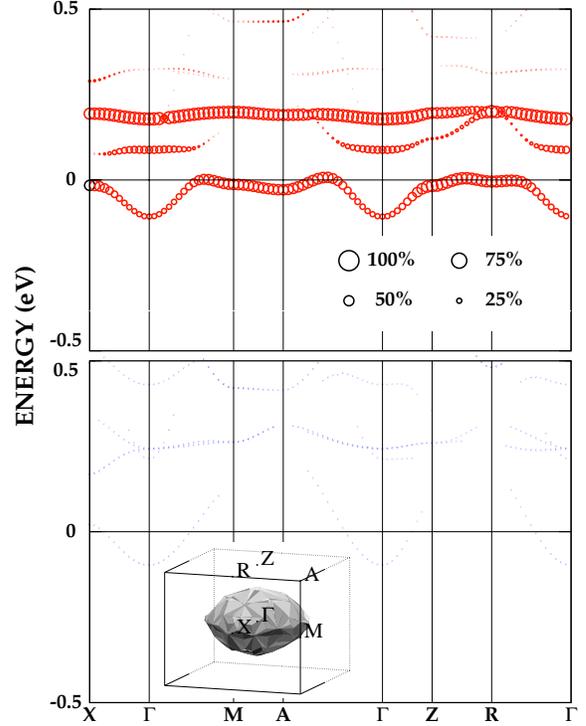,angle=270,width=8.0cm}}
\caption{Band structure of Li-intercalated 
$\rm Ti_{0.9375}Mn_{0.0625}O_{2}$ (Li/Ti=0.067) near the Fermi level.
The size of circle represents the amount of Mn 3$d$ component 
in the wave function
(the upper panel for the majority and the lower for the minority spin band). 
In the inset of the lower panel, the Fermi surface for the
minority spin band is given.
}
\label{Band}
\end{figure}

Now let us study the Li intercalation effects in 
$\rm Ti_{0.9375}Mn_{0.0625}O_{2}$.
Figure~\ref{MnLi1} provides the LSDA density of states (DOS)
for the FM phase of $\rm Ti_{0.9375}Mn_{0.0625}O_{2}$ with Li/Ti=0.067
\cite{Tote}. 
We have obtained metallic ground state, contrary to 
the non-intercalated Mn-doped TiO$_2$ which has an insulating ground state.
The energy band gap between O 2$p$ and Ti 3$d$ states
is estimated to be $\sim$ 3.2 eV,
which is reduced from $\sim$ 4 eV estimated for the non-intercalated case.
The Mn 3$d$ projected local density of states (PLDOS) 
indicates the high-spin state of Mn, that is, 
the exchange splitting between $t_{2g}$ states is larger than the 
crystal-field splitting between $t_{2g}$ and $e_g$ states.
Most of occupied Mn $3d$ ($t_{2g}$) states are located in the energy gap 
region, whereas the empty Mn $3d$ states are shifted up and
hybridized with Ti 3$d$ conduction band.
Note that, for the non-intercalated case, 
both the occupied $t_{2g}$ and the empty $e_g$ states were 
in the energy gap region far below the Ti 3$d$ conduction band.
According to the rigid band model,
the extra n-type carriers produced by Li intercalation  
would fill the empty Mn $e_g$ majority spin states first.
Figure~\ref{MnLi1}, however, indicates that the extra electrons 
fill not only Mn $e_g$ majority spin 
states but also Ti 3$d$ conduction band,
indicating that the simple rigid band concept does not work here.
It happens because the extra carriers produced by
Li intercalation are localized at Ti sites near 
the intercalated Li. Due to this localization,
Mn $e_g$ majority spin states are pushed up above the Ti 3$d$
conduction band bottom and so only slightly filled. 
The magnetic moment of Mn ion is 3.43 $\mu_B$, and the total magnetic 
moment is 3.74 $\mu_B$.
With Li intercalation,  magnetic moments of Ti ions are a bit enhanced
as compared to the  non-intercalated case.

Figure~\ref{Band} shows the band structure of 
Li-intercalated $\rm Ti_{0.9375}Mn_{0.0625}O_{2}$ 
with Li/Ti=0.067 near the Fermi level $\rm E_F$.
The size of circle represents the amount of Mn 3$d$ component in the 
wave function. It is seen that the majority spin states near $\rm E_F$
are mostly of Mn 3$d$ states,
while the minority spin states near $\rm E_F$ correspond mainly to
Ti 3$d$ conduction band states with fairly large band dispersion.
Thus Mn 3$d$ states are nearly half-metallic and
the flat Mn 3$d$ states suggest that they are rather 
localized\cite{Ldau}. 
Half-metallic Mn 3$d$ states near $\rm E_F$ together with the localized 
Mn $t_{2g}$ states far below $\rm E_F$ suggest that the double-exchange (DE) 
mechanism would be operative to yield the FM ordering of Mn spins\cite{Mspark}.

On the other hand,
the localized nature of Mn 3$d$ states and the dispersive 
Ti 3$d$ conduction band carriers are reminiscent of 
Mn-doped GaAs in which As $4p$ hole
carriers seem to mediate the Ruderman-Kittel-Kasuya-Yosida (RKKY)-type 
exchange interaction\cite{Shirai,Akai,JHPark}.
The difference from Mn-doped GaAs is the type of conduction band carriers
which is $n$-type in the present case.
With the information of the Fermi surface wave vector $k_F$ of the 
conduction band,
one can derive the magnetic ground state of the system in the RKKY formalism, 
That is, if the relation 2$k_FR$$\rm_{MM} < 4.5$ ($R$$\rm_{MM}$:
nearest distance between Mn spins) is satisfied,
the FM ordering of Mn spins would be stabilized\cite{RK}.
As shown in the inset of Fig.~\ref{Band}, the minority spin band structure 
has an oblate sphere-like Fermi surface centered at $\Gamma$.
One can measure $k_F$'s directed from $\Gamma$ to some
symmetry points in the irreducible Brillouin zone\cite{Fermi}.
The longest $k_F$ is $k_F^{\Gamma - X}$, while the shortest 
one is $k_F^{\Gamma - Z}$.
Then, with $R$$\rm_{MM}$= 7.57 \AA, one obtains 2$k_FR$$\rm_{MM} =$ 
4.63 and 2.96 for $\Gamma$ - X and $\Gamma$ - Z directions, respectively.
However, besides $k_F^{\Gamma - X}$, all other $k_F$'s give rise to
2$k_FR$$\rm_{MM}$ less than 4.5, suggesting that the FM interaction 
is dominant in the RKKY interaction mediated by the $n$-type 
conduction band carriers of Ti 3$d$ states.
Therefore, it is expected that,
in Li-intercalated $\rm Ti_{0.9375}Mn_{0.0625}O_{2}$,
the FM interaction between Mn spins is reinforced
by the combined effects of the DE and RKKY interactions.

For more Li-intercalated $\rm Ti_{0.9375}Mn_{0.0625}O_{2}$ with  Li/Ti=0.133,
we have also obtained magnetic and metallic ground state. 
More extra electrons fill both Mn and Ti 3$d$ states. 
Compared to the Li/Ti=0.067 case, the total magnetic moment
is reduced to 3.6 $\mu_B$, while Mn has nearly the same value, 3.41 $\mu_B$. 
The reduced total magnetic moment originates from lowered Mn 3$d$ DOS at
$\rm E_F$ which diminishes the spin-polarized hybridization 
between Mn and neighboring atoms.
As mentioned above, the observed FM state in Co-doped TiO$_2$
is questioned whether it is an intrinsic DMS property or not.
Once the FM state is observed in Mn-doped TiO$_2$,
it can be considered as intrinsic,
because Mn or Mn-oxide clusters, if they were formed in
anatase TiO$_2$, would have the AFM ground state.
Therefore, to get the direct evidence of DMS  property in TiO$_2$, 
it is desirable to fabricate Mn-doped TiO$_2$ with
varying the intercalated Li concentration.

\begin{figure}[t]
\centerline{
\epsfig{file=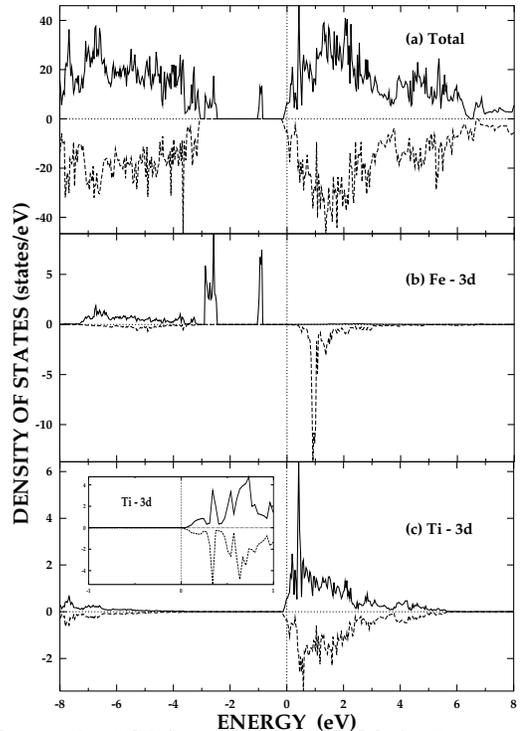,width=7.0cm}}
\caption{The LSDA total and PLDOS
   for Li-intercalated $\rm Ti_{0.9375}Fe_{0.0625}O_{2}$ with Li/Ti=0.133.
 Inset shows the insulating nature of Ti $3d$ PLDOS near $\rm E_F$ 
for Li/Ti=0.067.
}
\label{FeLi2}
\end{figure}

Next we have examined electronic structures of Li-intercalated
Fe-doped $\rm TiO_{2}$: $\rm Ti_{0.9375}Fe_{0.0625}O_{2}$ 
with Li/Ti=0.067 and 0.133.
For Li/Ti=0.067 case, we have obtained the insulating 
and magnetic ground state (see the inset of Fig.\ref{FeLi2}).
This is again in contrast to non-intercalated Fe-doped TiO$_2$ 
which is half-metallic\cite{Mspark}.
The electron carriers produced by Li intercalation fill up the 
Fe $e_g$ majority spin states which was half-filled for
the non-intercalated case. 
The empty Fe 3$d$ minority spin states are
shifted up above Ti $3d$ conduction band bottom,
as in the Mn-doped case (Fig.\ref{FeLi2}). 
The energy band gap between O $2p$ and Ti $3d$ states
is estimated to be $\sim$ 3.0 eV which is similar to 
that of Mn-doped case.
Total magnetic moment is 5 $\mu_B$, which comes mainly from Fe 
local magnetic moment of 4.31 $\mu_B$.  

Figure~\ref{FeLi2} presents results for more 
Li-intercalated Fe-doped TiO$_2$ with Li/Ti=0.133. 
In this case, we have obtained the metallic and magnetic ground state. 
The metallic ground state results from more electron carriers
produced by more intercalated Li concentration.
The states near $\rm E_F$ are mostly of Ti 3$d$ character.
This feature is different from the Mn-doped case having both Ti and
Mn 3$d$ characters near $\rm E_F$.
It is expected that the RKKY-type interaction works here too, 
that is, $n$-type carriers of Ti 3$d$ states mediate the
magnetic interaction between Fe local magnetic moments.
Note, however, that the Ti conduction band shows little spin polarization
at $\rm E_F$ ($\sim 20.5 \%$), and so this case is different from Mn-doped 
GaAs which has the 100$\%$ spin polarization at $\rm E_F$.
Total magnetic moment is a bit reduced to 4.85 $\mu_B$, but
Fe local moment 4.31 $\mu_B$ is nearly the same.
The reduced total
magnetic moment seems due to metallic phase, which 
results in the enhanced AFM interaction between Fe and
oxygen ions.

\begin{figure}[t]
\centerline{
\epsfig{file=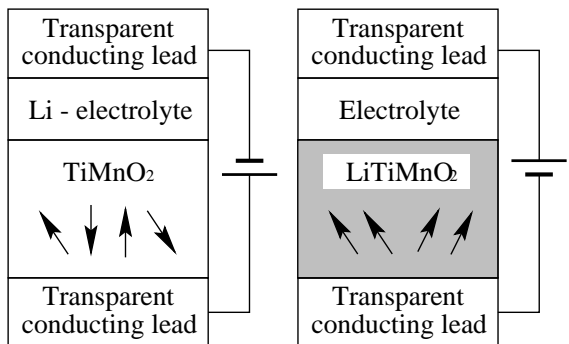,width=7.50cm}}
\caption{Schematic diagrams for an electric-field controlled nonvolatile FM
and electrochromic multifunctional device 
}
\label{Schem}
\end{figure}

On the basis of above findings, we propose a new spintronic device 
which can be controlled by the electric-field.
Figure~\ref{Schem} shows a schematic diagram of device made 
of Mn-doped TiO$_2$ operating by the voltage bias.
Li ions are supplied from the Li electrolyte.
The left figure represents the case of no Li intercalation
with applied (+) voltage bias on the bottom lead. Then it corresponds to 
non-intercalated Mn-doped TiO$_2$ which would not have the long-range
magnetic ordering (nonmagnetic: NM).
The right figure represents the case of Li intercalation
with applied ($-$) voltage bias on the bottom lead. 
Now, due to carriers produced by
Li intercalation, Mn-doped TiO$_2$ film would have the FM ordering.
In this way, it is possible to control the magnetic ordering 
by the electric-field.  
Even after removing bias, the magnetic ordering is maintained,
unless the reversed bias is applied.
So the proposed device has the non-volatile magnetic ordering.
The non-volatile property of this device is distinct from 
the previously demonstrated Mn-doped III-V device also controlled by
the electric-field\cite{Ohno1}. 
In addition, if the optical property is not affected by doped TM elements
as in the case of Co-doped TiO$_2$\cite{Matsumoto}, one can 
fabricate the multi-functional device
with variable magnetic and electrochromic properties:
\begin{eqnarray*}
Ti_{1-x}Co_xO_2 + yLi^++ye^- & \rightleftharpoons & Li_yTi_{1-x}Co_xO_2 \\
{\rm (transparent \ FM)} & & {\rm (colored \ NM)} \\
Ti_{1-x}\left(\begin{array}{l} Mn \\ Fe \end{array} \right)_xO_2 + 
yLi^++ye^- & \rightleftharpoons & Li_yTi_{1-x}\left(\begin{array}{l} Mn 
\\ Fe \end{array} \right)_xO_2 \\
{\rm (transparent \ NM)} & & {\rm (colored \ FM)}
\end{eqnarray*}
In this way,  an electric-field controlled spintronic
and electrochromic device can be realized in the
Li-intercalated TM-doped anatase TiO$_2$ system.

In conclusion, we have obtained ferromagnetic ground state for 
Li-intercalated Mn-doped TiO$_2$ with the $n$-type carriers
of Ti 3$d$ states.
Mn ions have the high spin states with 3.43 and 3.41 $\mu_{B}$ spin
moments for Li/Ti=0.067 and 0.133, respectively.
We have also obtained magnetic ground state for Li-intercalated Fe-doped 
TiO$_2$. As the intercalated Li concentration increases,
the electronic structure changes from insulating 
to metallic nature. Fe ions have the high spin states with 4.31 $\mu_{B}$ 
spin moment for both cases.
Our results suggest that the DMSs of Mn- and Fe-doped anatase TiO$_2$
can be synthesized by Li intercalation. 
Based on our findings, we propose a novel non-volatile
spintronic and electrochromic multifunctional device which can be
controlled by the electric-field.

Acknowledgments $-$
This work was supported by the KOSEF through the eSSC at POSTECH
and the Korea Research Foundation Grant (KRF-2002-070-C00038).


\begin{references}
\bibitem{Tote} 
We have compared the LSDA total energies of the FM and AFM 
configurations of TM (Mn, Fe) spins for Li-intercalated
 $\rm Ti_{0.875}TM_{0.125}O_{2}$ with Li/Ti=0.143. 
The energy differences between two configurations are found to 
be negligible within the numerical precision. But, for Fe-doped case,
the insulating state prefers the AFM configuration of Fe spins.
\bibitem{Ldau} The LSDA+$U$ ($U=3.0$ eV) band calculation yields the
further splitting in the majority spin $e_g$ states near $\rm E_F$.
But the lower pinned states at $\rm E_F$ are not changed much.
\bibitem{Fermi} 
We have checked that $k_F$'s are almost the same for
both FM and AFM configurations of Mn spins.
It is because $k_F$'s are determined by dispersive Ti 3$d$ states.
\end{references}
\end{document}